\documentclass[12pt,a4paper]{article}
\usepackage{amsmath}
\usepackage{amsfonts}
\usepackage{amssymb}
\usepackage{a4wide}
\newcommand{\initiate}{\setcounter{equation}{0}}
\def\Dirac{{\raise0.09em\hbox{/}}\kern-0.69em D}
 
\def\kbar{{\mathchar'26\mkern-9muk}}            
\def\lesssim{\mathrel{\hbox{\rlap{\hbox{\lower4pt\hbox{$\sim$}}}\hbox{$<$}}}}

\def\p{\partial}                                

\def\k{\kern-.1em\mathbin{,}\kern-.1em}
\def\hk{\kern.12em\raise-1em\hbox{$\hat{\raise1em\hbox{,}}$}\kern.12em}

\def\Sch{Schwarzschild}
\def\CA{{\cal A}} \def\CB{{\cal B}}
\def\CC{{\cal C}} \def\CD{{\cal D}}
\def\CE{{\cal E}} \def\CF{{\cal F}}
\def\CM{{\cal M}}

\begin{document}
\title{ Spherically Symmetric Noncommutative Space:\\ d\,=\,4}
\vskip25pt
\author{
        M. Buri\'c              $^{1}$\thanks{majab@phy.bg.ac.yu}
\ \
        J. Madore               $^{2}$\thanks{madore@th.u-psud.fr}
\\[25pt]
        $\strut^{1}$Faculty of Physics
\\               University of Belgrade, P.O. Box  368
\\               SR-11001 Belgrade
\\[10pt]
        $\strut^{2}$Laboratoire de Physique Th\'eorique
\\               Universit\'e de Paris-Sud, B\^atiment 211
\\               F-91405 Orsay
\\
}
\date{}
\maketitle

\parskip 10pt plus2pt minus2pt
\parindent 0pt

\vskip25pt

\abstract{In order to find a noncommutative analog of \Sch\  or \Sch-de Sitter black hole we investigate spherically symmetric spaces generated by four noncommutative coordinates in the frame formalism. We present two  solutions which however do not posess the prescribed commutative limit. Our analysis indicates that the appropriate noncommutative space might be found as a subspace of a higher-dimensional space.}

\thispagestyle{empty}

\vfill

 \newpage
\tableofcontents
\thispagestyle{empty} \newpage

\setlength{\parskip}{5pt plus2pt minus2pt}

\initiate
\section{Introduction}                   \label{I}

Proposals for a definition of noncommutative gravity are numerous and they vary in their physical motivation that is, basical starting point, and in their mathematical structure. For recent reviews, see~\cite{Fol,Sz}.
Obviously, one of the most important steps for a successful theory is to  describe spherically symmetric configurations.  This holds not only for  generalizations of  \Sch,   \Sch-de Sitter or Reissner-N\"ordstrom black holes but maybe more importantly for the cosmological solutions of noncommutative gravity. 

Noncommutative corrections to the black hole solutions  in the literature have been obtained using different approaches. In papers \cite{NSS} the authors analyze effects of noncommutativity by assuming that the source is not of  a $\delta$-function form (as in Einstein gravity for the \Sch\  black hole) but an origin-centered Gaussian distribution. The noncommutativity is `integrated out' and effective; the  gravity  is described classically. A similar assumption is made in~\cite{Kob} where the dominant contribution to the deformation of space-time comes from the noncommutative scalar field. In papers~\cite{Chai,Muk} the deformation to the black hole geometry is due to the noncommutativity of space-time itself; corrections are found both to geometric and to thermodynamic quantities. The framework which was
used is that of~\cite{Cham}; as in~\cite{Wess}, gravity is defined as a gauge field corresponding to the invariance under diffeomorphisms, while the tetrad and the spin connection are represented as fields in the Moyal-deformed space-time. This means that, regarding the algebraic structure, the `ground state' is a space with constant noncommutativity of Cartesian coordinates, 
\begin{equation}
[x^\mu,x^\nu] = i\theta^{\mu\nu} = i\kbar J^{\mu\nu}={\rm const}.
                                           \label{flat}
\end{equation} 
(We introduce here paralelly our notation: the $\kbar$ is a dimensional constant which measures noncommutativity, the commutator is denoted by $J^{\mu\nu}$ as we keep the letter $\theta^\alpha$ for the tetrad.) 
Noncommutative BTZ black holes were discussed in~\cite{BTZ}.

It is clear that the assumption (\ref{flat}) breaks rotational invariance, and thus  the  corrections found using it are not spherically symmetric. For this reason the flat noncommutative space (\ref{flat}) is  not well suited for a  study of spherical symmetry. On the other hand, noncommutative spherically symmetric spaces surely must exist. Breaking of this symmetry, even if averaged over smaller regions, certainly would have an impact on the structure of the universe at large scales as well as observable consequences.

The approach we advocate as a natural one and close to the intuition of  a physicist is that of the moving frames~\cite{BOOK}. Since one of its first results~\cite{Mad92a} was to describe and explain the geometry of the fuzzy sphere one would expect that a generalization to the four-dimensionsional space  would be  easy. To formulate an algebra of noncommuting coordinates with  appropriate symmetries is indeed not difficult and has been done in a couple of variants~\cite{Mou,Schupp,tobe}. If however we wish to impose the conditions of differentiability and of a prescribed commutative limit, then the task is indeed more difficult. In this paper we will present some examples of spaces with spherical symmetry based on four-dimensional algebras with differential calculi defined by moving frames. The organization of the paper is as follows: in Section~2 we introduce the main notions of the noncommutative frame formalism. In Section~3 we analyze in some detail the simplest example of a spherically symmetric four-dimensional space, while in Section~4 we give a generalization. The drawbacks of the given models are discussed in the concluding section.

\initiate
\section{Formalism}

A simple and computationally very efficient way to describe Einstein gravity is the moving frame formalism of Cartan~\cite{Car}. Geometry of a specific space can be fully described by its moving frame or Vielbein, which is a set of 1-forms $\theta^\alpha$,
\begin{equation}
\theta^\alpha = \theta^\alpha_\mu (x)\, dx^\mu ,
\end{equation} 
and their dual vector fields $e_\alpha$,
\begin{equation}
  \theta^\alpha(e_\beta) =\delta^\alpha_\beta .
\end{equation} 
The moving frame is  preferred to other bases of 1-forms because the inverse metric calculated in this basis is a constant matrix
\begin{equation}
g(\theta^\alpha\otimes\theta^\beta) = g^{\alpha\beta} ={\rm const}. \label{g}
\end{equation} 
Often one takes the Minkowski values $g^{\alpha\beta} =\eta^{\alpha\beta}$ and then the tangent vectors $e_\alpha$ 
are orthonormal. If the space posesses symmetries, the frame is adapted to them.

The differential of an arbitrary function $f$ can be written as
\begin{equation}
df =(e_\alpha f) \, \theta^\alpha ,                 \label{d}
\end{equation} 
in particular
\begin{equation}
dx^\mu = (e_\alpha x^\mu)\, \theta^\alpha = e^\mu_\alpha(x)\, \theta^\alpha.
\end{equation} 
Differential of the frame forms defines the Ricci rotation coefficients $C^\alpha{}_{\beta\gamma}$,
\begin{equation}
d\theta^\alpha = -\frac 12 C^\alpha{}_{\beta\gamma} \theta^\beta \theta^\gamma .
\end{equation} 
From these, assuming that the space is without torsion, one  obtains the components of the connection 1-form
$ \omega^\alpha{}_\beta = \omega^\alpha{}_{\gamma\beta}\theta^\gamma $\ 
as
\begin{equation}
\omega_{\alpha\gamma\beta} = - \frac 12( C_{\alpha\beta\gamma} - C_{\gamma\alpha\beta} + C_{\beta\gamma\alpha}) ;                     \label{omega}
\end{equation} 
the connection defines the curvature 
\begin{equation}
\Omega^\alpha{}_\beta = d\omega^\alpha{}_\beta + \omega^\alpha{}_\gamma \omega^\gamma{}_\beta ,                                          \label{Omega}
\end{equation} 
and this, along with~(\ref{g}), concludes the geometric characterization of the given space. 

It is possible as we shall see to describe geometry of a noncommutative space in a similar manner. Here we  just outline the main idea and give some  details which we need later on; a rigorous definition, further properties and more examples  can be found in \cite{BOOK,Mou95,DubMadMasMou96b,BurGraMadZou06a,BurMad05b}. A noncommutative space is an algebra ${\cal A}$ generated by a set of noncommuting hermitean coordinates $x^\mu$ which satisfy a commutation relation of a general form
\begin{equation}
[x^\mu,x^\nu] = i\kbar J^{\mu\nu}(x) .             \label{J}
\end{equation} 
In order to define the differential structure on $\CA$ in analogy with~(\ref{d}) one has to specify a noncommutative frame $\theta^\alpha$.  Condition (\ref{g}) that the components $g^{\alpha\beta}$ of the metric  are constant means that they commute with all elements of the algebra, that is with its generators,
\begin{equation}
[ g^{\alpha\beta}, x^\mu ] = 0.
\end{equation} 
Sufficient to insure this property is to assume that the frame 1-forms themselves commute with $x^\mu$
\begin{equation}
[\theta^\alpha, x^\mu] = 0 ,                                     \label{thetax}
\end{equation} 
and this is, apart from the linearity of the metric, one of the main inputs of the construction. Having the differential defined and imposing the Leibniz rules one can proceed and define the connection, the torsion and the curvature. 

The frame can be given either by 1-forms $\theta^\alpha$ or by the dual derivations $e_\alpha$. A derivation can be inner, defined in terms of the elements of the algebra
\begin{equation}
e_\alpha f = [p_\alpha, f] .
\end{equation} 
Operators $p_\alpha$ we then call the momenta; specification of all momenta gives the frame. Imposing the condition $d^2 =0$ on the differential, one obtains\footnote{If, in addition, the assumptions that the momenta generate $\CA$ and that the only central element is identity are made.}~\cite{BOOK}
 that the momentum algebra cannot be arbitrary as the position algebra~(\ref{J}): it is quadratic,
\begin{equation}
[p_\alpha,p_\beta] = (i\kbar)^{-1} K_{\alpha\beta} + F^\gamma {}_{\alpha\beta} p_\gamma -i\kbar Q^{\gamma\delta}{}_{\alpha\beta} p_\gamma p_\delta .
                                        \label{pp}
\end{equation} 
 The case when  momenta are in the algebra
is in a way  typical for noncommutative spaces: for example, for finite dimensional matrix algebras all derivations are inner. In the flat noncommutative space defined by relation~(\ref{flat}), when
 $J^{\mu\nu}$ is nonsingular the momenta are given by
\begin{equation}
p_\alpha = (i\kbar)^{-1}\delta^\mu_\alpha J^{-1}_{\mu\nu} x^\nu .
\end{equation} 
In principle however one need not restrict to 
inner derivations.
In the ordinary quantum-mechanical situation, where the space is flat and commutative, the momenta are outer derivations
\begin{equation}
p_\alpha = (i\hbar)^{-1} \delta^\mu _\alpha \, \frac{\p}{\p x^\mu} .
\end{equation} 

To recapitulate: we start with the algebra
\begin{equation}
[x^\mu,x^\nu] = i\kbar J^{\mu\nu}(x)  ,              
\end{equation} 
and with the differential $d$,
\begin{equation}
dx^\mu = e^\mu_\alpha(x) \theta^\alpha.
\end{equation} 
The algebra is associative, so the commutators obey the Jacobi identities
\begin{equation}
[x^\lambda, J^{\mu\nu}] + [x^\nu, J^{\lambda\mu}] + [x^\mu, J^{\nu\lambda}] = 0;
                                                      \label{one}
\end{equation} 
the differential obeys the Leibniz rule
\begin{equation}
d(fh) = df\, h + f\, dh .
\end{equation} 
Consistency of the  algebraic and the  differential structures gives the  condition
\begin{equation}
[e^\mu_\alpha, x^\nu] + [x^\mu, e^\nu _\alpha] = i\kbar e_\alpha J^{\mu\nu} ,
                                                     \label{two}
\end{equation} 
obtained by differentiating~(\ref{J}). 
A further restriction on the algebra, obtained from~(\ref{thetax}), is
\begin{equation}
[x^\mu, C^{[\alpha}{}_{\gamma\delta}]\theta^{\beta]}_\mu = 0 . \label{three}
\end{equation} 
Thus a noncommutative geometry can be specified by a set of functions $J^{\mu\nu}(x)$ and $e^\mu_\alpha(x)$ which satisfy relations (\ref{one}-\ref{three}). 
These are the equations which we will try to solve under the additional, specific assumptions about the symmetries. We will not impose any further restrictions in the form of field equations on the metric  as, having not specified the representation of the algebra, we cannot speak of the action or of the action principle. We will assume that
the connection and the curvature are given by formulae~(\ref{omega}-\ref{Omega}). The metric by linearity is
\begin{equation}
g^{\mu\nu}(x) = g\big(\theta^\mu_\alpha(x) \theta^\alpha \otimes \theta^\nu_\beta(x)\theta^\beta\big) = \theta^\mu_\alpha(x) \theta^\nu_\beta(x) \eta^{\alpha\beta}.
\end{equation} 
Obviously, the commutative limit  is automatically given as the limit of functions $\theta^\alpha_\mu (x)$ when their arguments $x^\mu$ commute, that is when $\kbar \to 0$.

Solving the commutator equations within an abstract algebra is a  difficult task and therefore we choose to work in the `semiclassical' that is almost commutative approximation. We  assume that the parameter $\kbar$ is small and thus we solve the equations only in the leading order in $\kbar$. The main formula which we use is
\begin{equation}
[x^\mu,f] = i\kbar J^{\mu\nu}\p_\nu f + o(\kbar^2).
\end{equation} 
It is easy to check its validity; also, it is easy to see that the notion of partial derivative $\p_\nu f$ which we  use is meaningful in this approximation.

\initiate
\section{Example}

We realize the spherical symmetry by an appropriate choice of the Ansatz for the frame. We use the isotropic coordinates $x^i = x, y, z$ ($i = 1,2,3$): 
\begin{equation}
r^2 = x^2 + y^2 + z^2 = (x^i)^2,
\end{equation} 
and $x^0 = t$. The coordinates are fixed here; the
 formalism is to first order covariant under the change of coordinates, as discussed in~\cite{BurGraMadZou06a}. 
 For the sake of custom still we will  keep the (apparent) covariance: the spatial coordinate indices $i,j$ we will  lower with the Cronecker delta 
$\delta^i_j$ when the summation convention is used. Moreover, instead of  coordinate indices $i,j$ we will sometimes use  frame indices $a,b$ for coordinates, shortening thus $ \delta^a_i x^i = x^a $.
 
The Ansatz for the commutators is
\begin{eqnarray}
&&[t, x^i] = i\kbar J^{0i} = i\kbar \gamma(r,t) x^i   \label{tx}\\[4pt]
&& [x^i, x^j] = i\kbar J^{ij} = i\kbar \epsilon^{ij}{}_{k}\, \alpha(r,t) x^k .
                                                    \label{xx}
\end{eqnarray} 
We have therefore
\begin{equation}
[r^2, x^i] = -i\kbar\epsilon^{i}{}_{jk} (x^j \alpha x^k +\alpha x^k x^j ) = o(\kbar^2) ,
\end{equation} 
so to first order
\begin{equation}
[r,x^i] = 0  .            \label{rx}
\end{equation} 
We will assume that (\ref{rx}) is exact. From
(\ref{tx}) we obtain
\begin{equation}
[t,r] = i\kbar \beta = i\kbar \gamma r .
\end{equation} 
The Jacobi identities restrict functions $\gamma$ and $\alpha$. From
\begin{equation}
[t,[x^i,x^j]] + [x^j, [t, x^i]] + [x^i, [x^j, t]] = 0
\end{equation} 
to  leading order we obtain
\begin{equation}
\frac{\p \alpha}{\p r}\, \gamma r = \alpha \gamma   .   \label{txx}
\end{equation} 
The other Jacobi identity
\begin{equation}
\epsilon_{ijk}[x^i,[x^j,x^k]] =0 ,
\end{equation} 
gives
\begin{equation}
\gamma \frac{\p \alpha}{\p t} = 0.                      \label{xxx}
\end{equation} 
The simplest solution to these two equations is $\gamma =0$, in which case the matrix $J^{\mu\nu}$ is degenerate and $J^{0i} =0$. We will  not consider this possibility in any detail, 
 as it is basically  an extension of the fuzzy sphere by a commutative time coordinate. 

For $\gamma \neq 0$ the solution to the given equations is $\alpha = A r$. We explore first the simplest possibility, $A=0$ that is $J^{ij} =0$, which is also degenerate. Furthermore, in order to get a static metric, we assume that $\gamma$ depends only on radius $r$. As we shall see the algebra consistent with this Ansatz is that of the $\kappa$-Minkowski space.
We start with 
\begin{eqnarray}
&& [t,x^i] = i\kbar \gamma x^i  ,    \label{t,x}\\[4pt]
&& [x^i,x^j] =0      ,               \label{x,x}\\[4pt]
&&[t,r] = i\kbar \gamma r   ,         \label{t,r}
\end{eqnarray} 
$\gamma = \gamma(r)$.
To define  differential calculus we need a frame: we assume
\begin{equation}
dt = \CA \theta^0,\qquad dx^i = \CF\delta^i_a\theta^a ,    \label{frame}
\end{equation} 
and  $\CA = \CA(r)$, $\CF = \CF(r)$.

The Leibniz constraints can be obtained by differentiating commutators. The only nontrivial equation is
\begin{equation}
[dt, x^i] + [t,dx^i] = i\kbar (\gamma dx^i + \gamma^\prime dr x^i) ,
                                               \label{339}
\end{equation} 
where we have denoted $\gamma^\prime = \frac{\p \gamma}{\p r}$.
Since to first order we can write
\begin{equation}
dr = r^{-1} x_l dx^l = r^{-1} \CF x_a \theta^a ,
\end{equation} 
on the right hand side we have
\begin{equation}
\gamma dx^i + \gamma^\prime dr\,  x^i =(\gamma \delta^{i}_{j} + \gamma^\prime r^{-1} x^i x_j) dx^j .
\end{equation} 
On the other hand, from the commutation of $\theta^\alpha$ with the algebra we obtain that
\begin{equation}
[dt, x^i] = [\CA\theta^0,x^i] =0,
\end{equation} 
and therefore (\ref{339}) reduces to
\begin{equation}
[t,\CF] \theta^a = i\kbar \CF(\gamma  \delta^a_b + \gamma^\prime r^{-1} x^a x_b) \theta^b .                  \label{1}
\end{equation} 
The solution to~(\ref{1}) is $\gamma^\prime =0$, $\CF = \mu r$. As $\gamma$ is constant, this is the $\kappa$-Minkowski space, $\kappa =\kbar\gamma$.

There is a further nontrivial restriction which comes from  the stability of  the condition $[t,\theta^0] =0$ under differentiation. Using $\theta^0 = \CA^{-1} dt$ we have
\begin{equation}
d\theta^0 = -r^{-1} \CA^\prime \CA^{-1} \CF x_a \theta^a \theta^0 =
- \mu \CA^\prime \CA^{-1} x_a \theta^a \theta^0 . 
\end{equation} 
Therefore the equation
\begin{equation}
d[t,\theta^0] = [dt, \theta^0] + [t, d\theta^0] = 0
\end{equation} 
gives the constraint
\begin{equation}
\frac{\CA^\prime}{\CA}+(\frac{\CA^\prime}{\CA})^\prime r = 0, \label{**}
\end{equation} 
because, from the general arguments~\cite{BOOK}, we know that the coefficients in the anticommutator $[\theta^\alpha, \theta^\beta]$ have to be constant and equal to the corresponding ones in~(\ref{pp}),
\begin{equation}
[\theta^\alpha, \theta^\beta] = 2 i\kbar Q^{\alpha\beta}{}_{\gamma\delta}\theta^\gamma \theta^\delta.
\end{equation} 
Equation (\ref{**}) has a solution $\CA = (\mu r)^n$ for arbitrary power 
$n$. In particular, the frame is of the  \Sch\  form for $n=1$.

The $\kappa$-Minkowski space allows  other four-dimensional frames. 
Introducing the change of coordinates  $y^i = \log x^i$, relations 
\begin{equation}
[t,x^i] = i\kbar \gamma x^i ,\qquad [x^i, x^j] =0,
\end{equation} 
can be rewritten as 
\begin{equation}
[t,y^i] = i\kbar \gamma  ,\qquad [y^i, y^j] =0.
\end{equation}
 In coordinates $t$, $y^i$ the noncommutativity is constant and one can consistently choose another  frame, the flat one:
\begin{equation}
{\theta^{ 0}}^\prime = dt, \qquad {\theta^{ a}}^\prime = \delta^a_idy^i = (x^a)^{-1} dx^a \ \ ({\rm no\ summation}).
\end{equation} 
It is not suprising that the same algebra can support different geometries.

Going back to the original example: the limiting commutative line element which corresponds to~(\ref{frame}) is
\begin{eqnarray}
&& ds^2 = -(\theta^0)^2 + (\theta^a)^2 = -(\mu r)^{2n} dt^2 + \frac{1}{(\mu r)^2} \, (dx^i)^2 \nonumber \\[4pt]
&&\phantom{ds^2 } = -(\mu r)^{2n} dt^2 + \frac{1}{(\mu r)^2}\, dr^2 +  \frac{1}{\mu^2 }\,  d\Omega^2 .        \label{metric}
\end{eqnarray} 
This space is  spherically symmetric and curved: its the scalar curvature
is constant, $R=2(1-n^2)\mu^2$.
For the particular value $n = 1$ it has an interesting property: the corresponding Einstein tensor is equal to the energy-momentum tensor of the electromagnetic field defined by
\begin{equation}
F_{\mu\nu} = (-g)^{\frac{1}{4}} J_{\mu\nu} .
\end{equation} 
In this particular case the elecrtic field is nonzero
while the magnetic field vanishes:
\begin{equation}
E_i = F_{0i}=\frac{\gamma}{\mu r}\, x^i ,\qquad
B^i = \frac{1}{2}\epsilon^{ijk}F_{jk} = 0 ,
\end{equation} 
and it is easy to check that
\begin{equation}
R_{\mu\nu} -\frac{1}{2}g_{\mu\nu} R = -16\pi G_N \big(F_{\mu\rho}F_\mu{}^\rho - \frac{1}{4}g_{\mu\nu}F_{\rho\sigma} F^{\rho\sigma}\big)
\end{equation} 
for an appropriate choice of $\gamma$.
This supports the idea presented before that noncommutativity of the space-time can be interpreted as an additional  source of gravity (`Poisson energy'). Apart from a spin-1 it also has a scalar mode, \cite{BurMadZou07}.

The main drawback of the given example, except for the absence of the Newtonian or  SdS limit, is that metrically the space   is a direct product of $(t,r)$ and $(\theta,\phi)$ subspaces. The angular part $d\Omega^2$ is multiplied by a constant in (\ref{metric}) instead by the $r^2$.  Therefore one can not introduce, by any change of coordinates, a radius corresponding to the surface of the sphere. Except through the commutation relations, the factor spaces are unrelated.

\initiate
\section{Generalization}

One could think that degeneracy of the four-dimensional space
discussed in the previous section is
 due to degeneracy of the commutators and that a choice
\begin{eqnarray}
&& [t,x^i] = i\kbar \gamma x^i   ,                   \\[4pt]
&& [x^i,x^j] = i\kbar \epsilon^{ij}{}_{k}  r x^k ,        \label{commutators}   \\[4pt]
&&[t,r] = i\kbar \gamma(r) r      ,                 
\end{eqnarray} 
might be a better starting point. 
However it is easy to see that the frame
\begin{equation}
dt = \CA \theta^0,\qquad dx^i = \CF\delta^i_a \theta^a
\end{equation} 
 makes~(\ref{commutators}) inconsistent with the Leibniz rules. Namely, from
\begin{equation}
[dx^i, x^j] + [x^i, dx^j] = i\kbar \epsilon^{ij}{}_{k} (dr \, x^k + r dx^k) ,
\end{equation} 
immediately follows that  $\CF = 0$. Therefore we choose
 the most general Ansatz within the given context.
We assume 
\begin{equation}
J^{0i} = \gamma(r,t)\, x^i, \qquad J^{ij} = \epsilon^{ij}{}_{k} r x^k.
\end{equation} 
The matrix $J^{\mu\nu}$ is now invertible:
\begin{equation}
J^{-1}_{0i} = -\frac{1}{r^2\gamma }\, x_i, \qquad J^{-1}_{ij} = -\frac{1}{r^3\,}\epsilon_{ijk} x^k ,
\end{equation} 
and therefore we can expect that the momenta are inside the algebra. 
For the frame  we take
\begin{eqnarray}
&& dx^0 = \CA \theta^0 + \CB x_a \theta^a , \label{A1}\\[4pt]
&& dx^m = \CC x^m \theta^0 + (\CF \delta^m_a + \CD x^m x_a + \CE \epsilon ^{m}{}_{ab} x^b )\theta^a ,                     \label{A2}
\end{eqnarray} 
that is 
\begin{eqnarray}
&& e^0_0 = \CA , \\[4pt]
&& e^0_i = \CB x_i  , \\[4pt]
&& e^m_0 =   \CC x^m    ,      \\[4pt]
&& e^m_i = \CF \delta^m_i + \CD x^m x_i + \CE \epsilon ^{m}{}_{ij} x^j .
\end{eqnarray} 
We allow the dependence on $t$ and $r$ for
 functions $\CA$, $\CB$, $\CC$, $\CD$, $\CE$, $\CF$.

In order to obtain relations among the given functions we impose the
Leibniz rules.
The first one,
\begin{equation}
[dt, x^i] + [t, dx^i] = i\kbar d(\gamma x^i),
\end{equation} 
projected to the frame basis gives two equations:
\begin{eqnarray}
&& [e^0_0,x^i] +[t, e^i_0] = i\kbar \big( e^0_0 \p_0(\gamma x^i) + e^k_0 \p_k(\gamma x^i) \big) , \\[4pt]
&& [e^0_m,x^i] +[t, e^i_m] = i\kbar \big( e^0_m \p_0(\gamma x^i) + e^k_m \p_k(\gamma x^i) \big) .
\end{eqnarray} 
The first of these two can be easily simplified:
\begin{equation}
\dot \CA \gamma - \CA \dot\gamma = r(\CC \gamma^\prime - \CC^\prime \gamma),
                                   \label{*}                 
\end{equation} 
with $\dot \CA = \p_0 \CA$, $\CA^\prime =\p_r \CA$. Other Leibniz rules require more work. Analyzing all of them we obtain the following set of equations
\begin{eqnarray}
&& \dot \CA \gamma - \CA \dot\gamma = r(\CC \gamma^\prime - \CC^\prime \gamma),
\\[4pt]  &&
r^2(\dot\CB\gamma -\CB \dot\gamma +\CD^\prime \gamma r -\CD \gamma^\prime r +\CD \gamma) = -3\CF^\prime \gamma r+\CF\gamma^\prime r+ 3\CF \gamma,
\\[4pt]  &&
-\CB r +\CE^\prime \gamma r =0,
\\[4pt]  &&
\CD r +\dot \CE \gamma = -\frac{\CF}{r},
\\[4pt]  &&
\CF =0 .
\end{eqnarray} 
Their solution is
\begin{equation}
\CF =0,\qquad \CB = \CE^\prime \gamma,\qquad \CD = -\dot \CE \, \frac{\gamma}{r} .
\end{equation} 
  $\CA$, $\CC$ and $\gamma$ are related by~(\ref{*}), otherwise arbitrary.

The previous equations can be also solved for the momenta. The  $p_0$ is defined by
\begin{eqnarray}
&& [p_0, t ] = e^0_0 = \CA , \\[4pt]
&& [p_0, x^i] = e^i_0 = \CC x^i .
\end{eqnarray} 
The upper equations can be rewritten as
\begin{equation}
 \frac{\p p_0}{\p r} = -\frac{ \CA}{\gamma r}, \qquad \frac{\p p_0}{\p t} = \frac{\CC}{\gamma} ,
\end{equation} 
 their integrability condition is~(\ref{*}). They have as
solution
\begin{equation}
i\kbar p_0 = - \int \frac{\CA}{\gamma r}dr + \int \frac{\CC}{\gamma}dt = \CM .
\end{equation} 
For spatial momenta $p_i$ the equations  are
\begin{eqnarray}
&& [p_a, t] = e^0_a = \CB x_a , \\[4pt]
&& [p_a, x^m ] = e^m_a = \CD x^m x_a +\CE \epsilon^{m}{}_{aj} x^j ,
\end{eqnarray} 
 the solution is
\begin{equation}
i\kbar p_a = \frac{\CE}{r}\, x_a .
\end{equation} 

Let us discuss the commutative limit of (\ref{A1}-\ref{A2}).  To obtain the line element 
\begin{equation}
ds^2 = -(\theta^0)^2 +(\theta^i)^2
\end{equation} 
 we need to invert the $e^\mu_\alpha$. It is easy to check that to first order the inverse matrix $\theta^\alpha_\mu$ is given by
\begin{equation}
\theta^\alpha_\mu = \frac{1}{\Delta}
\left(\begin{array}{cc}
\CD  \ \ & -\CB r^{-2}x_m \\[6pt]
-\CC r^{-2}x^a  \ \  & \CA r^{-4}x_m x^a - \Delta (\CE r^2)^{-1}\epsilon^{a}{}_{mn} x^n
\end{array} \right)
\end{equation} 
with 
\begin{equation}
\Delta = \CA \CD - \CB \CC = -\gamma \,(\frac{\CA}{r}\dot \CE + \CC \CE^\prime ).
\end{equation} 
We have therefore 
\begin{eqnarray}
&& \theta^0 = \frac{\CD}{\Delta}dt - \frac{\CB}{r^2 \Delta}x_i dx^i = \frac{\CD}{\Delta}dt - \frac{\CB}{r \Delta} dr , \\[6pt]
&& \theta^a = -\frac{\CC}{r^2\Delta}x^a dt +\frac{\CA}{r^3\Delta}x^a dr - \frac{1}{r^2 \CE}\epsilon^{a}{}_{mn}x^m dx^n .
\end{eqnarray} 
For the line element we obtain
\begin{equation}
ds^2 = -(\frac{\CD}{\Delta}dt -\frac{\CB}{r\Delta}dr )^2 + (\frac{\CC}{r\Delta}dt -\frac{\CA}{r^2\Delta}dr)^2 + \frac{1}{\CE^2}\, d\Omega^2 .
\end{equation} 
However we see that, changing  coordinates $(t,r)$ to $(\CM,\CE)$, $ds^2$ can be written as
\begin{equation}
ds^2 = \frac{\gamma^2}{r^2 \Delta^2}(- d\CE ^2 + d\CM ^2) + \frac{1}{\CE^2}\, d\Omega^2 .
\end{equation} 
To first order
\begin{equation}
[\CE,\CM] = i\kbar \Gamma =\frac{\p(\CE,\CM )}{\p( t,r)}\, i\kbar \gamma r = i\kbar \frac{r\Delta}{\gamma} ,
\end{equation} 
so we have finally
\begin{equation}
ds^2 = \frac{1}{\Gamma^2}(- d\CE ^2 + d\CM ^2) + \frac{1}{\CE^2}d\Omega^2 .
                                            \label{line}
\end{equation} 

\initiate
\section{Conclusion}

The covariance of the formalism under the change of coordinates has reduced a seemingly general Ansatz (\ref{A1}-\ref{A2}) to a simple form (\ref{line}). As  the commutator $\Gamma$ is {\it a priori} an unconstrained function we will not calculate the curvature;
however a couple of observations are in order. It is interesting to note that the metric is the simplest in coordinates which are related to the momenta. In these `natural' coordinates the role of the radius has $\CE$, which enters~(\ref{line}) with the  signature `-'. The \Sch\ metric behaves similarly  inside the horizon where  the radius and the time change roles, except that here we have that the function $\Gamma^2$ does not change the sign. This shows that the solution is not static. Were the signature Euclidean, the obtained solution would be easier to interpret. This could mean that we need more than three spatial dimensions to obtain a proper \Sch-like commutative limit.

The space which we have just described has another problem: the momentum algebra. Calculating the commutator
\begin{equation}
[p_a,p_b] = \epsilon_{ab}{}^{c} \CE p_c
\end{equation} 
we see that the algebra is not quadratic in the momenta: that is, unless we add one more momentum, for example $p_r = \CE$. 
Then we would need also to add an independent coordinate, $r$. The relation $r^2 = (x^i)^2  $ would then be interpreted as a constraint  defining a four-dimensional  subspace of a five-dimensional space.
In any case we can conclude that the Ansatz~(\ref{A1}-\ref{A2}) is overconstraining and that a way of relaxing it is necessary. We will return to this issue in a forthcoming publication~\cite{tobe}.

The presented examples also show  advantages of the frame formalism. It presents a well defined and consistent procedure to formulate noncommutative gravity.  In its spirit the formalism is geometric; as we have seen, it is both covariant under transformation of coordinates and  adapted to  description of symmetries. A further property is that it is defined in a representation-free manner, which means that it describes equally well spaces with infinite
and finite-dimensional representations. However as we have seen, in its simplest or most straightforward realizations it proves to be somewhat rigid; it seems it should be applied `with soul'.

\vskip0.5cm
\noindent{\bf Acknowledgement}  \ This work was supported  by Grant 2091 of the  Quantum Geometry and Quantum
Gravity Network (ESF) and by Grant 141036  of the Serbian Ministry of Science.


\providecommand{\href}[2]{#2}\begingroup\raggedright\endgroup

\end{document}